\documentclass[a4paper]{spie}  

\usepackage{amsmath,amsfonts,amssymb}
\usepackage{graphicx}
\usepackage{upgreek}
\usepackage[colorlinks=true, allcolors=blue]{hyperref}

\newcommand*\aap{A\&A}

\newcommand*\ao{Appl Opt}

\newcommand*\apjl{ApJ}

\newcommand*\apjs{ApJS}

\usepackage{amsmath,amsfonts,amssymb}
\usepackage{graphicx}
\usepackage[colorlinks=true, allcolors=blue]{hyperref}
\usepackage{threeparttable} 
\usepackage{macro_ref}
\usepackage{geometry}
\title{L-band nulling interferometry at the VLTI with Asgard/NOTT: status and plans} 

\author[a]{Denis Defrère}
\author[a]{Romain Laugier}
\author[a]{Marc-Antoine Martinod}
\author[a]{Germain Garreau}
\author[a]{Kwinten Missiaen}
\author[a]{Muhammad Salman}
\author[a]{Gert Raskin}
\author[b]{Colin Dandumont}
\author[c,d]{Steve Ertel}
\author[e]{Michael J. Ireland}
\author[f]{Stefan Kraus}
\author[g]{Lucas Labadie}
\author[b]{Alexandra Mazzoli}
\author[h]{Gyorgy Medgyesi}
\author[g,i]{Ahmed Sanny}
\author[b]{Olivier Absil}
\author[j]{Peter Ábráham}
\author[k]{Jean-Philippe Berger}
\author[k]{Myriam Bonduelle}
\author[a]{Azzurra Bigioli}
\author[l]{Emilie Bouzerand}
\author[e]{Josh Carter}
\author[m]{Nick Cvetojevic}
\author[e]{Benjamin Courtney-Barrer}
\author[l]{Adrian M. Glauser}
\author[i]{Simon Gross}
\author[n]{Xavier Haubois}
\author[a]{Noel James}
\author[j]{Andras Peter Joo}
\author[m]{Stephane Lagarde}
\author[o]{Alain L\'eger}
\author[c]{Jarron Leisenring}
\author[p]{Jérôme Loicq}
\author[k]{Guillermo Martin}
\author[m]{Frantz Martinache}
\author[j]{Gyorgy Mezo}
\author[m]{Sébastien Morel}
\author[a]{Johan Morren}
\author[o]{Marc Ollivier}
\author[q]{Gordon Robertson}
\author[b,c]{Hélène Rousseau}
\author[e]{Warrick Schofield}
\author[n]{Nicolas Schuhler}
\author[q]{Adam Taras}
\author[a]{Bart Vandenbussche}
\author[n]{Julien Woillez}
\affil[$a$]{Institute of Astronomy, KU Leuven, Celestijnenlaan 200D, 3001, Leuven, Belgium}
\affil[$b$]{Space sciences, Technologies \& Astrophysics Research (STAR) Institute, University of Li\`ege, Li\`ege, Belgium}
\affil[$c$]{Department of Astronomy and Steward Observatory, The University of Arizona, 933 North Cherry Ave, Tucson, AZ 89 85721, USA}
\affil[$d$]{Large Binocular Telescope Observatory, The University of Arizona, 933 North Cherry Ave, Tucson, AZ 85721, USA}
\affil[$e$]{Research School of Astronomy and Astrophysics, Australian National University, Canberra, ACT 2611, Australia}
\affil[$f$]{School of Physics and Astronomy, University of Exeter, Exeter, United Kingdom}
\affil[$g$]{I. Physikalisches Institut, Universit\"at zu K\"oln, Z\"ulpicher Str. 77, 50937 Cologne, Germany}
\affil[$h$]{Fornax, H-1123 Budapest, Táltos u. 1., Hungary}
\affil[$i$]{MQ Photonics Research Centre, School of Mathematical and Physical Sciences, Macquarie University, NSW, 2109, Australia}
\affil[$j$]{Konkoly Observatory, HUN-REN Research Centre for Astronomy and Earth Sciences, Budapest, Hungary}
\affil[$k$]{University Grenoble Alpes, CNRS, IPAG, 38000 Grenoble, France}
\affil[$l$]{Eidgen\"ossische Technische Hochschule (ETH) Zurich, Institute for Particle Physics and Astrophysics, Zurich, Switzerland}
\affil[$m$]{Laboratoire Lagrange, Universit\'e C\^ote d'Azur, Observatoire de la C\^ote d'Azur, CNRS, Boulevard de l'Observatoire, CS 34229, 06304, Nice, France}
\affil[$n$]{European Southern Observatory, Karl-Schwarzschild-Straße 2, 85748 Garching, Germany}
\affil[$o$]{Institut d'Astrophysique Spatiale, Universit\'e Paris Sud}
\affil[$p$]{Delft University of Technology, Netherlands}
\affil[$q$]{Sydney Institute for Astronomy, School of Physics, University of Sydney, NSW 2006, Australia}

\authorinfo{Further author information: Denis Defr\`ere: E-mail: denis.defrere@kuleuven.be}

\begin{document} 
\maketitle

\begin{abstract}
NOTT (formerly Hi-5) is the L’-band (3.5-4.0~$\mu$m) nulling interferometer of Asgard, an instrument suite in preparation for the VLTI visitor focus. The primary scientific objectives of NOTT include characterizing (i) young planetary systems near the snow line, a critical region for giant planet formation, and (ii) nearby main-sequence stars close to the habitable zone, with a focus on detecting exozodiacal dust that could obscure Earth-like planets. In 2023-2024, the final warm optics have been procured and assembled in a new laboratory at KU Leuven. First fringes and null measurements were obtained using a Gallium Lanthanum Sulfide (GLS) photonic chip that was also tested at cryogenic temperatures. In this paper, we present an overall update of the NOTT project with a particular focus on the cold mechanical design, the first results in the laboratory with the final NOTT warm optics, and the ongoing Asgard integration activities. We also report on other ongoing activities such as the characterization of the photonic chip (GLS, LiNbO$_3$, SiO), the development of the exoplanet science case, the design of the dispersion control module, and the progress with the self-calibration data reduction software. 
\end{abstract}


\section{Introduction}
\label{sec:intro}  

Asgard/NOTT\cite{defrere_path_2018,defrere_hi-5_2018,defrere_l-band_2022} is the L-band nuller of Asgard, an instrument suite officially recommended in June 2023 by ESO's Scientific Technical Committee (STC) for the visitor focus of the VLTI \cite{martinod_high-angular_2023}. Asgard/NOTT follows a series of nulling interferometers\cite{bracewell_detecting_1978} that have been deployed over the past three decades on state-of-the-art facilities, both across single telescopes and as separate aperture interferometers\cite{spalding_unveiling_2022}. These include the BracewelL Infrared Nulling Cryostat\cite{hinz_imaging_1998}, the Keck Interferometer Nuller\cite{serabyn_keck_2012}, the Palomar Fiber Nuller\cite{mennesson_high-contrast_2011,serabyn_nulling_2019}, the Large Binocular Telescope Interferometer\cite{hinz_overview_2016,ertel_overview_2020}, and DRAGONFLY/GLINT on Subaru/SCExAO\cite{norris_first_2020}. 

Leveraging the experience gained with these nulling instruments, Asgard/NOTT will be the first nulling interferometer for the \emph{Very Large Telescope Interferometer} (VLTI). Thanks to the VLTI state-of-the-art infrastructure, long baselines, and recent GRAVITY+ upgrade, Asgard/NOTT has the potential to carry out several exoplanet programmes to study young Jupiter-like exoplanets at the most relevant angular separations\cite{fernandes_hints_2019,fulton_california_2021} (i.e., close to the snow line) and better understand how planets form and evolve. In 2023-2024, the warm optical design has been finalized\cite{garreau_asgardnott_2024} and all warm components besides the dispersion corrector have been procured. The dispersion corrector has been designed and presented elsewhere in these proceedings (see Laugier et al.). The instrument assembly in a new laboratory at KU Leuven and first null measurements are presented in a companion paper in these proceedings (Garreau et al.). In parallel, the photonic beam combiner has been manufactured and tested using three different technologies and/or materials: GLS (see Sanny et al. in these proceedings), LiNbO$_3$ (see Bonduelle et al. in these proceedings), and SiO manufactured by the VTT company. The designs of the cold mechanical components and cryostat have progressed towards maturity and the first images from the HAWAII-2RG detector have been obtained using the MACIE electronics. The software has been further developed, both a flexible Python-based scripting approach for engineering purposes and an ESO-compliant framework in collaboration with the Asgard consortium. 

In addition to the technical and laboratory activities, progress has been made on both the exoplanet science case based on new end-to-end performance simulations\cite{laugier_asgardnott_2023} and the nulling self-calibration pipeline (presented separately in these proceedings, see Martinod et al.) The NOTT project received funding from OPTICON for a preliminary study\cite{defrere_hi-5_2018} (2016-2020) and was later funded by the European Research Council (2020-2025). In this paper, we give an update of the project, its science case and current status.

\section{NOTT science case}
\label{sec:science}

\begin{figure}[t]
\centering
\includegraphics[width=0.47\textwidth]{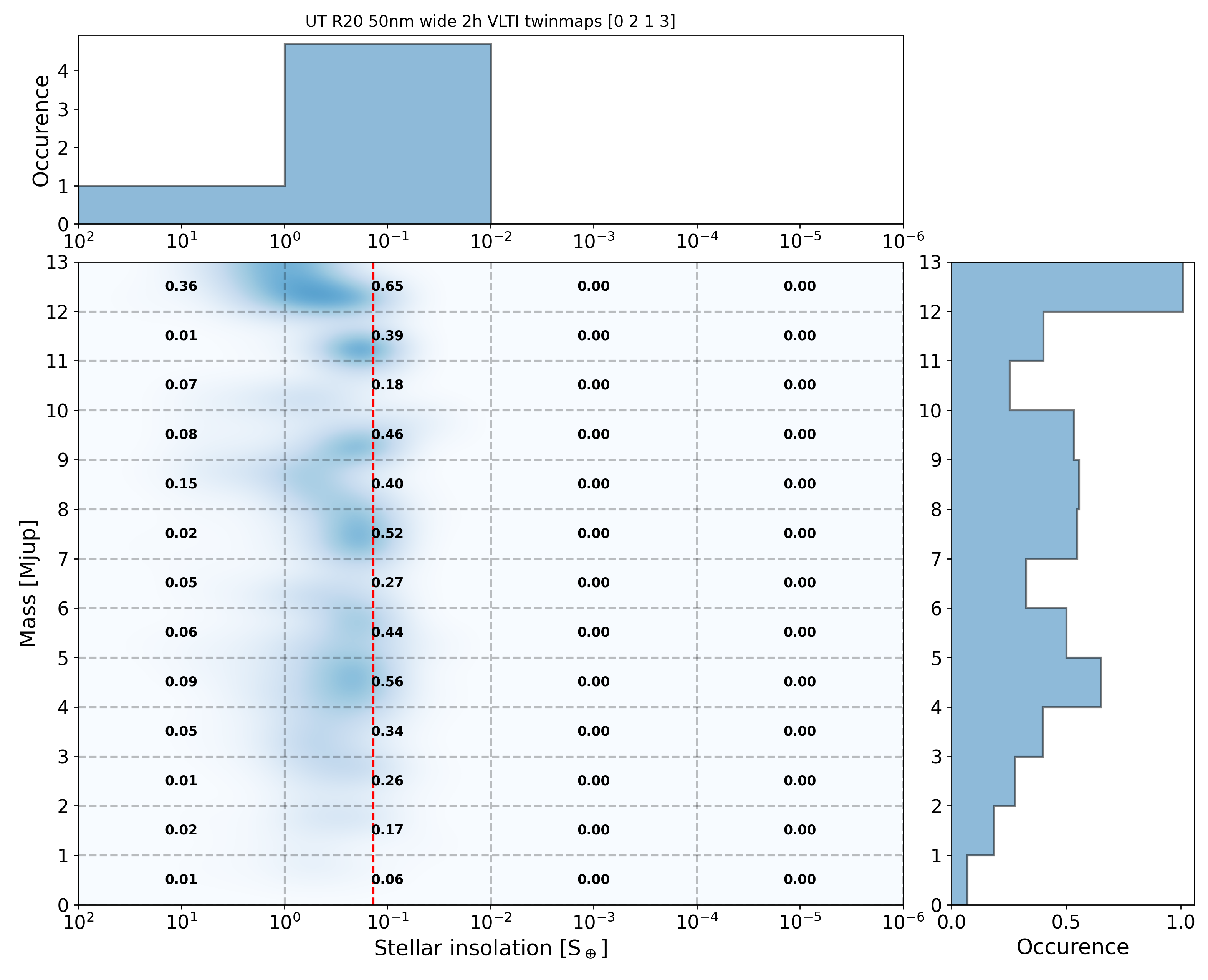}
\includegraphics[width=0.50\textwidth]{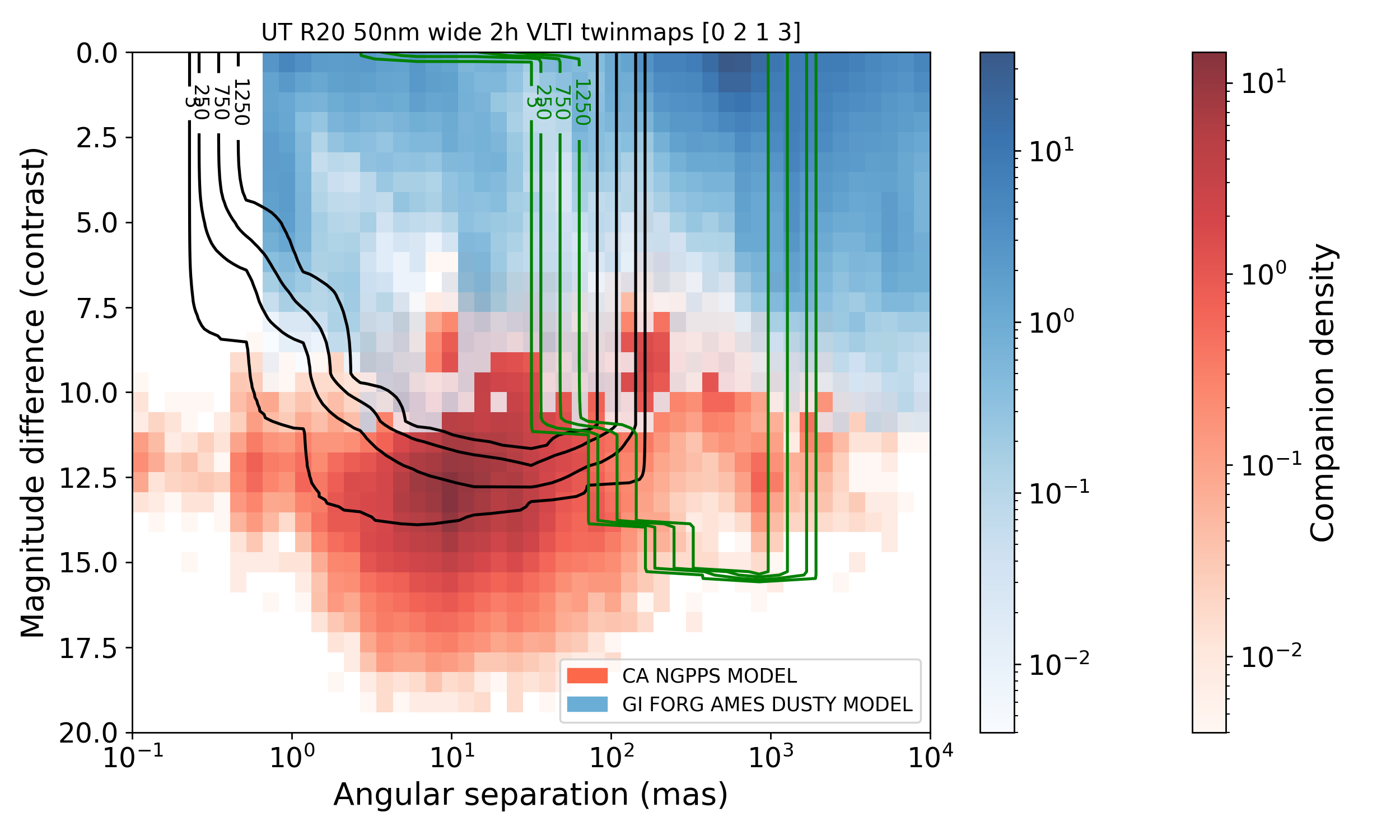}
\caption{\emph{Left}: Mass-stellar insolation range for the CA-detected planets with the UTs around all stars in the catalog (integration time of 2h per star). The number in each dotted square represents the corresponding planet's occurrence rate. The top and left histograms are the integrated values (sum over the axis). The red dotted line represented the snow line region. \emph{Right}: NOTT depth of search for the UTs (black curves), compared to the one from SHINE\cite{vigan_sphere_2020} (green curves). The contour lines give the number of stars around which the instrument is sensitive as a function of contrast (differential magnitude between the host star and the planet) and semi-major axis. The planetary population comprises two formation scenarios. Their probability density is represented, from white to blue for the GI formation model, and from white to red for the CA formation model.}
\label{Fig:ContrastUT}
\end{figure}

The main NOTT science cases cover exozodiacal disks\cite{defrere_l-band_2022}, the characterization of hot Jupiters\cite{defrere_hi-5_2018}, and planet formation\cite{wallace_likelihood_2019,wallace_constraints_2021,dandumont_vltihi-5_2022}. In this paper, we only present an update of the exoplanet science case based on Dandumont et al.\ (in prep). We refer readers interested in exozodiacal dust to the pertinent literature\cite{ertel_near-infrared_2014,ertel_near-infrared_2016,absil_near-infrared_2021,ertel_hosts_2018,ertel_hosts_2020,kirchschlager_first_2020,pearce_hot_2022}. Using core accretion (CA) and a gravitational instability (GI) models, we generated putative exoplanets and brown dwarves around a sample of 1484 nearby ($<$150pc) young ($<$300Myr) stars\cite{gagne_banyan_2018} and computed their detectability with our end-to-end performance simulator\cite{laugier_asgardnott_2023}.  In short, this simulator computes sensitivity maps in the NOTT field of view for given star magnitudes and declinations, taking into account the covariance matrix of the errors and statistical detection tests (based on likelihood ratios). These sensitivity maps, which provide the maximum magnitude of detectable companions in the field of view, are compared to the planets generated with the planet formation models. Some examples can be found on the Github page\footnote{Nulling VLTI simulator: \url{https://github.com/rlaugier/SCIFYsim}}. The results are shown in Figure~\ref{Fig:ContrastUT} for the UTs. Thanks to the long VLTI baselines, Asgard/NOTT could observe exoplanets closer to their host stars than classical direct imaging instrument such as SPHERE. Considering all sub-stellar companions generated by the planet formation models, NOTT should be able to detect about four CA planets with the UTs. These planets are more massive than Jupiter and mostly located near the snow line of their host star (see Figure~\ref{Fig:ContrastUT}, left). For GI sub-stellar companions, NOTT could lead to the detection of around ten brown dwarves with the ATs. Future work will focus on defining and optimizing the observing strategy as well as updating the yield when GAIA DR4 become available. 

\section{Instrumental overview}
\subsection{Assembly and first fringes with warm optics}

The complete NOTT optical design was completed in 2022 and reviewed in June 2023 as part of an overall Asgard review. After a few revisions, the final design was published with a tolerance analysis\cite{garreau_design_2022, garreau_asgardnott_2024}.  The design is implemented to be highly symmetric, in order to avoid polarization or phase mismatch between the four arms. In 2023-2024, the final warm optics have been procured and assembled in a new laboratory on the ground floor at KU Leuven. The only missing component is the dispersion corrector, which is discussed separately in these proceedings (see Laugier et al.).

\begin{figure}[!t]
    \centering
    \includegraphics[width=0.95\textwidth]{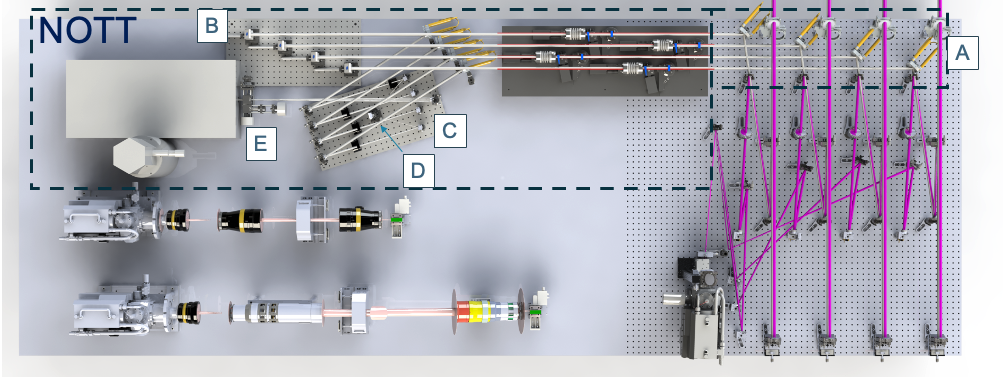}
    \includegraphics[width=0.95\textwidth]{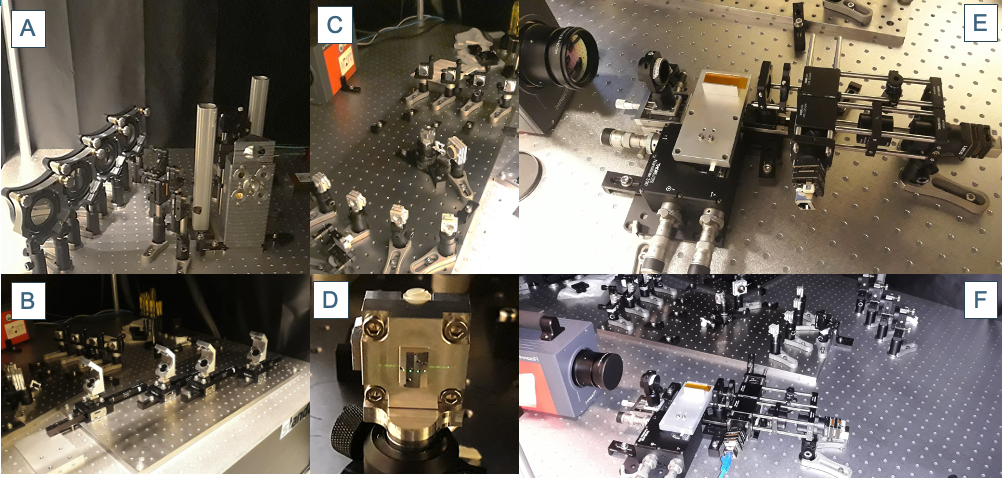}
    \caption{\emph{Top}: top view design of Asgard on the visitor table (only lower level), showing NOTT within the dashed blue lines.  After wavefront and pathlength correction with HEIMDALLR, the four beams coming from the VLTI (purple lines) are transmitted to the NOTT optics via CaF$_2$ dichroics from Rocky Mountain Instruments (see top right part). The beams are then reflected on tip/tilt mirrors (see A) along the edge of the optical table, reflected on the delay lines (see B), and directed to the injection system (see C). The injection system uses a mirror slicer (see D) from the Weylandts company. Finally, the four beams are injected into the photonic chip (see E). \emph{Bottom}: Corresponding pictures of the current laboratory implementation at KU Leuven. }
    \label{Fig:vlti_lab}
\end{figure}

Figure~\ref{Fig:vlti_lab} shows the top view of the Asgard visitor table (only lower level) and corresponding pictures taken in the laboratory at KU Leuven. The four VLTI beams (purple) first go through the HEIMDALLR module\cite{ireland_image-plane_2018, 
Taras:24} to perform fringe tracking and wavefront correction with adaptive optics. Four dichroics made of calcium fluoride (CaF$_2$) then transmit the L'-band toward the NOTT instrument. Mirrors are then used to direct the four beams along the edge of the optical table. There are two Tip/Tilt Mirrors (TTMs) and one delay line for each beam. Using four mirror slices, called ``slicer", and Off-Axis Paraboloids (OAPs), the system recombines the four pupils of the beams before the injection into the chip. The combined pupil plane is located at the same position as the aperture stop before the injection lens. The delay lines and Pupil Recombination Optics are both assembled on their own respective breadboards to facilitate their moving to Paranal. More information, including the complete list of optics and an overview of first NOTT fringes, can be found in a companion paper in these proceedings (Garreau et al.). 

\subsection{Mechanical design of the cold optics}

\begin{figure}[!t]
\begin{center}
\begin{tabular}{c}
\includegraphics[width=0.98\textwidth]{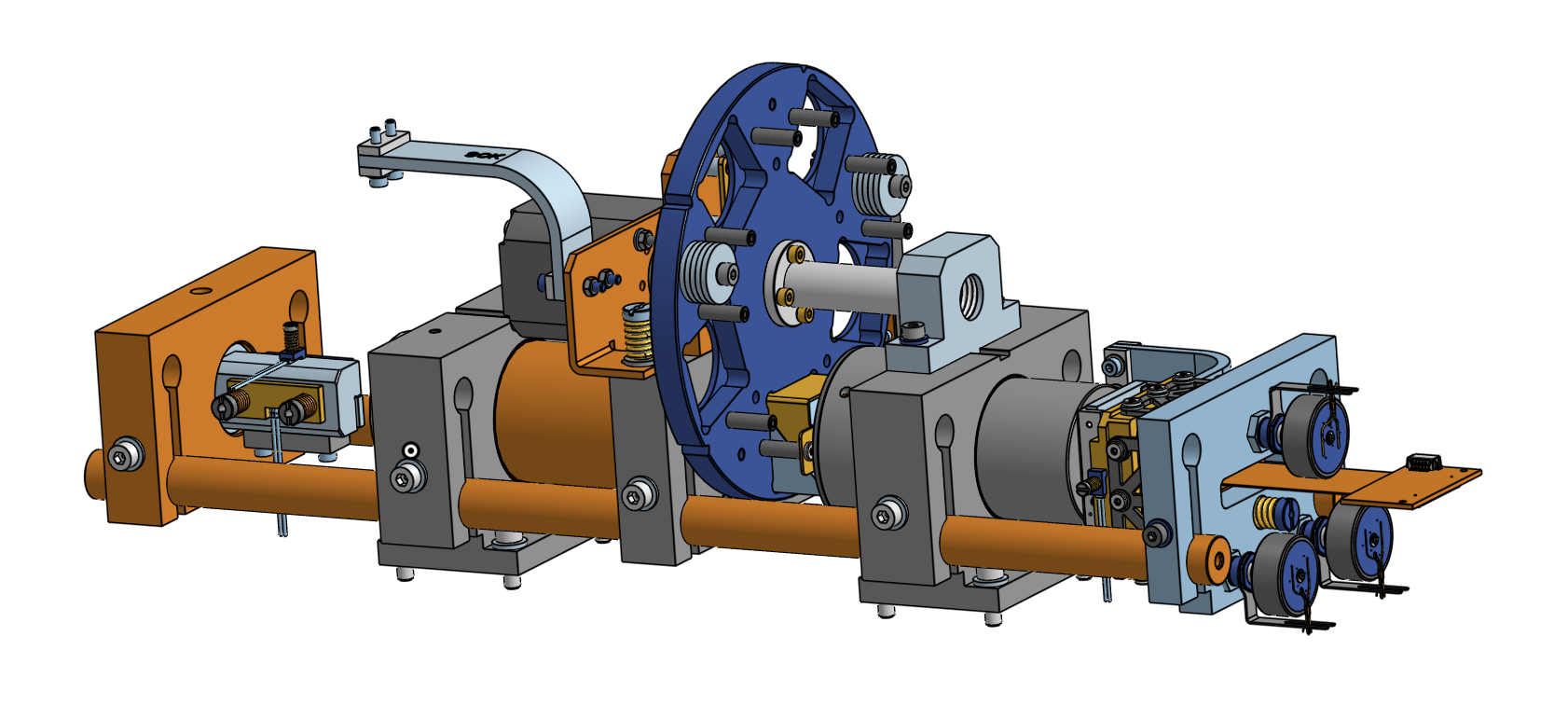}
\end{tabular}
\end{center}
\caption{\label{Fig:Fullspectro} Mechanical design of the NOTT mechanical assembly showing from left to right: the support of the photonic chip, the collimator and Wollaston support assembly, the filter wheel for the grisms and shutters, the imager assembly, the detector support mounted on a focusing stage.}
\end{figure}

The optical design and requirements of the NOTT spectrograph have been completed in 2022\cite{dandumont_technical_2022}. Here, we present an update of the mechanical design as shown in Figure~\ref{Fig:Fullspectro}. The mechanical structure includes:
\begin{itemize}
\item supporting structure, consisting of two holding clamps and two round horizontal bars;
\item the support of the photonic chip, which can be rotated by 180 degrees and is connected to a temperature sensor;
\item the collimator and Wollaston support assembly, which holds the three lenses of the collimator. The Wollaston prism is used to split the polarizations;
\item the filter wheel with grisms of different spectral resolutions and a cold shutter;
\item the imager assembly,  which holds the four lenses of the imager;
\item the focusing stage, which holds the the HAWAII-2RG\textregistered \ detector and is mounted on piezo translation stage;
\end{itemize}

All mechanical parts are made of aluminium. The screws, springs and washers are made of stainless steel. The supporting structure has a holding clamp which is near the focusing stage and is rigidly bolted down to the base plate; the other clamp and the focusing stage are held down to the base plate with disc springs. This enables some parts of the supporting structure to move relative to the base
plate if the coefficient of thermal expansion (CTE) of the material of the base plate and the material of the mechanical parts differ.

Each lens is held firmly against its seat with a flexible centering ring. Each centering ring is split and has a number of cutouts in order to make it more flexible. The centering rings are pressed against the beveled surfaces of the
lenses with small helical springs. This solution has the following advantages:
\begin{itemize}
\item The axial position of each lens is determined by the position of its flat seating surface
only. It is not sensitive to the manufacturing tolerances of either the lens or the
centering ring. Therefore the axial position of the lenses after cooling down to 90K
can be very accurately predicted.
\item Each lens is always perfectly centered because there is no gap between the lens and
the centering ring and there is no gap between the centering ring and the tube.
\item The lenses are never subjected to excessive forces either at room temperature or at
90K (or anywhere in between) because the centering rings can slide inside the tube
and compensate for dimensional changes. Only the spring preload will change
slightly.
\end{itemize}

Because all parts of the supporting structure are made of aluminium, the shrinkage of the aluminium parts from 293K to 90K must be calculated. A simple method would be to calculate with the coefficient of thermal expansion for aluminium (23.6$\times$10$^{-6}$ $C^{-1}$). This would give 23.6$\times$10$^{-6}$ $\times$ (293-90) = 0.00479 = 0.479\% contraction. However at cryogenic temperatures the coefficient of thermal expansion itself changes so a more accurate method is required. Using the linear expansion coefficient as a function of temperature, we get 0.38\% contraction from 293K to 90K. This means that the dimensions of the parts of the supporting structure must be 1.00381 times longer at room temperature if we want to maintain the correct distances between the lens elements at 90K.

\subsection{Software and control}

\begin{figure}[!t]
\begin{center}
\begin{tabular}{c}
\includegraphics[width=0.95\textwidth]{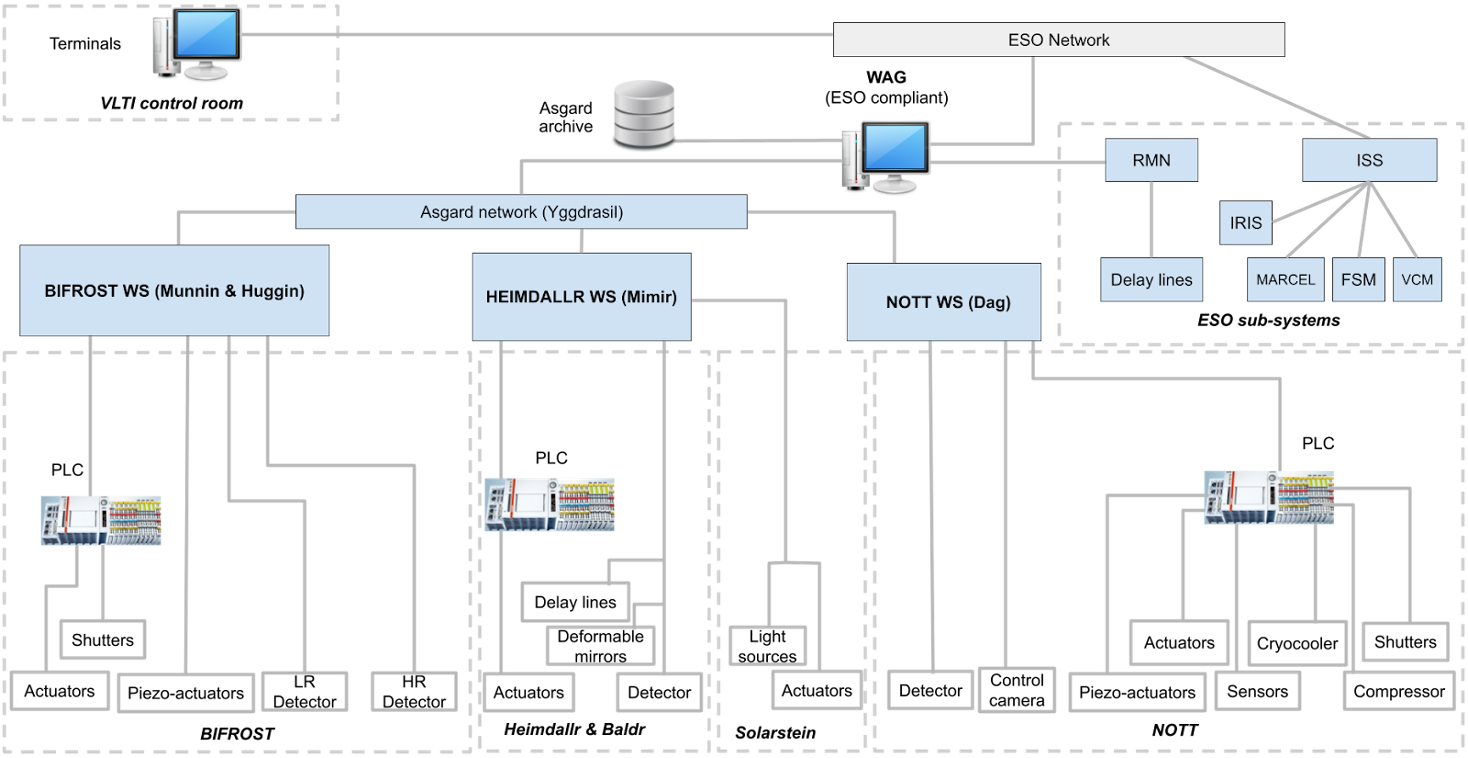}
\end{tabular}
\end{center}
\caption{\label{fig:asgard_architecture} Control architecture of Asgard. WAG is an ESO-compliant machine that can communicate with ESO facility (in particular the RMN and the ISS) and control every Asgard subsystem through their respective workstations. Each instrument has one or two workstations to control its optics and cameras via programmable logic controllers (PLC). The users operate Asgard from terminals located in the VLTI control room that will connect to the WAG workstation. NOTT is controlled by a R7515 Dell server running Linux called \emph{Dag} and two PLCs.}
\end{figure}

The overall Asgard control architecture is shown in Figure~\ref{fig:asgard_architecture}. The hardware architecture is designed so that all the instruments can be operated from the terminals in the VLTI control room. This centralisation allows the observer to run all the instruments simultaneously and to communicate with the VLTI systems. Asgard low-level IT hardware and software are non-ESO compliant, hence must be kept separate. In addition, they need to communicate with each other; hence, they are connected to each other on a server-client system. Asgard instruments need to interact with ESO systems such as the Reflective Memory Network (RMN), the delay lines and the calibration sources. The solution is to connect all instruments’ workstations, via a local network called \emph{Yggdrasil}, to an ESO compliant machine, the instrument Workstation of AsGard, (WAG, formerly Hermod workstation) which is on ESO’s network. WAG will interact with ESO systems, the RMN, and transmit data to and from Asgard's modules that require such data. WAG respects ESO’s standards for hardware and software. From the ESO network point of view, Asgard is operated from a single computer represented by WAG. Each instrument has one or two workstations called \emph{Module Control Units} (MCU). The NOTT module control unit, named Dag, controls the camera and the opto-mechanical devices of NOTT. It is a R7515 Dell server running Linux. Besides controlling some sub-systems directly, the workstations are also using \emph{Programmable Logic Controllers} (PLC). The choice of driving a sub-system directly from the workstation or a PLC is motivated by the latency, frequency of usage and the most convenient way to drive the component while respecting constraints such as interface or power consumption. In 2024, a first version of the ESO-compliant WAG software has been deployed on the NOTT workstation and is currently being tested. More information on the overall Asgard control architecture and software can be found in a companion paper in these proceedings (see Martinod et al.).

\subsection{Detector characterization}

\begin{figure}[!t]
\begin{center}
\begin{tabular}{c}
\includegraphics[width=0.95\textwidth]{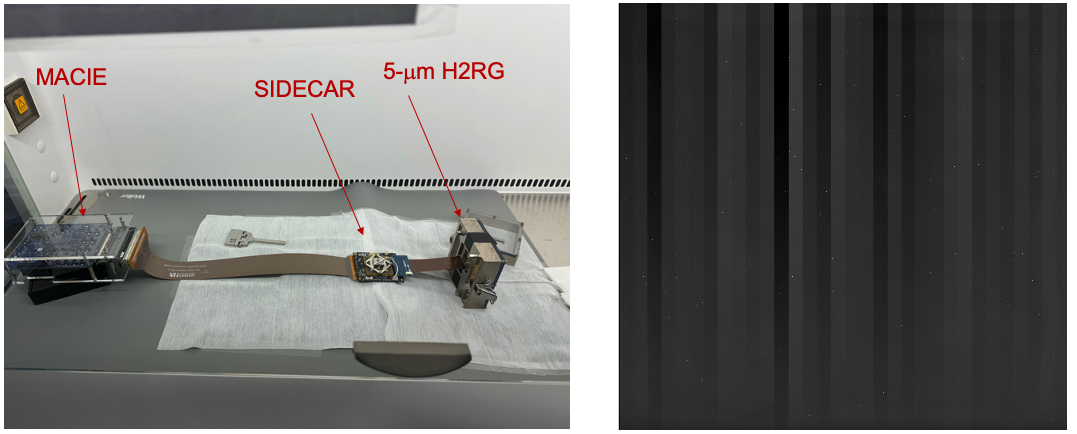}
\end{tabular}
\end{center}
\caption{\label{Fig:detector} \emph{Left:} Laboratory setup of the Teledyne’s 5-micron HAWAII-2RG\textregistered~detector with SIDECAR cold electronics and Markury Scientific MACIE warm electronics.\emph{Right:} First image obtained at warm temperature in slow mode.}
\end{figure}

During the winter 2023, we installed the 5-micron HAWAII-2RG (H2RG) detector in a cleaned lab space at KU Leuven for testing and characterization. The setup consists of Teledyne’s 5-micron H2RG with SIDECAR ASIC cold electronics controlled by Markury Scientific's MACIE warm electronics that can be accessed from the internal network. The setup and first image of the detector are shown in Figure~\ref{Fig:detector}. Current work consists in characterizing the detector cosmetics and implementing the different observing modes based on the software used in LBTI/LMIRCam\cite{leisenring_-sky_2012}. 

\section{Data reduction pipeline}

State-of-the-art nulling interferometers use the so-called nulling self-calibration (NSC) technique to achieve high-contrast interferometric nulling. This method has shown a gain in contrast by a factor 10 on the Palomar Fiber Nuller\cite{hanot_improving_2011,mennesson_new_2011,serabyn_nulling_2019}, the LBTI\cite{defrere_nulling_2016, mennesson_making_2016} and GLINT\cite{norris_first_2020, martinod_achromatic_2021}. The NSC approach is a simulation-based inference method. Monte-Carlo simulations with a simulator of the instrument are carried out to reproduce the measured distributions. This method generates sequences of quantities based on measured or assumed distributions of the sources of noise and the astrophysical null depth, and propagates them throughout the simulator to obtain a simulated sequence of nulled flux. The generated data are sorted into a histogram, which is then fit to the histogram of the real data to find the best-fit parameters. This process enables the self-calibration, providing the simulator can correctly model the dominant effects of the instrument response. By relying on a Monte-Carlo approach and a clearly defined instrumental function, this technique can be used on any nuller. In the context of the NOTT project, we have developed a Generic data Reduction for nulling Interferometry Package\footnote{Nulling self-calibration package: \url{https://github.com/mamartinod/grip}} (GRIP) based on the experience gained with existing nulling interferometers. More information can be found in a companion paper in these proceedings (Martinod et al.). In parallel, we also investigated the use of Principal Component analysis to improve background subtraction~\cite{rousseau_improving_2024}. Further work is currently being carried out develop a uniform nulling data standard\footnote{Nulling data standard: \url{https://github.com/rlaugier/nifits}}. 

\section{Summary and future work}

NOTT is the L’-band (3.5-4.0 $\mu$m) high-contrast imager of Asgard, an instrument suite in preparation for the visitor focus of the VLTI. Together with the other Asgard instruments, NOTT was reviewed by an internal expert panel in June 2023 and is currently under review by ESO staff for software and hardware interfaces, with targeted integration at Paranal in 2025. In this paper, we present an update on the project, including the first fringes and null measurements in the laboratory, the development of the cold mechanical design, the development of the ESO-compliant control software, the detector characterization, progress with the exoplanet science case, and a new generic data reduction pipeline. Short term work is now focused on integration activities, the finalization of the cryostat design, and the procurement of all cold mechanical and optical components.
 
\acknowledgments 
SCIFY has received funding from the European Research Council (ERC) under the European Union's Horizon 2020 research and innovation program (grant agreement CoG - 866070). This project has received funding from the European Union’s Horizon 2020 research and innovation programme under grant agreement No 101004719. S.K. acknowledges support from an ERC Consolidator Grant (``GAIA-BIFROST'', grant agreement No.\ 101003096). SE is supported by the National Aeronautics and Space Administration through the Astrophysics Decadal Survey Precursor Science program (Grant No. 80NSSC23K1473). We are grateful for the kind support and constructive interactions with colleagues at ESO, in particular Frédéric Gonte and Antoine Mérand.

\bibliographystyle{spiebib} 

\end{document}